\documentclass[epj,final]{svjour} 
\usepackage{graphics}
\usepackage{graphicx} 
\usepackage{color}
\def\lsim{\mathrel{\rlap{
\lower4pt\hbox{\hskip-3pt$\sim$}}
    \raise1pt\hbox{$<$}}}     
\def\gsim{\mathrel{\rlap{
\lower4pt\hbox{\hskip-3pt$\sim$}}
    \raise1pt\hbox{$>$}}}     
\begin{document}

\title{Estimation of the Shear Viscosity from 3FD Simulations of Au+Au Collisions 
at $\sqrt{s_{NN}}=$ 3.3--39 GeV}
\titlerunning{Estimation of the shear viscosity}
\author{Yu. B. Ivanov\inst{1,2}  \thanks{e-mail: Y.Ivanov@gsi.de} \and
A. A. Soldatov\inst{2} \thanks{e-mail: saa@ru.net}}
\institute{%
National Research Centre "Kurchatov Institute", 
123182 Moscow, Russia
\and
National Research Nuclear University "MEPhI" (Moscow Engineering
Physics Institute), 115409 Moscow, Russia}
\date{Received: date / Revised version: date}

\abstract{
An effective shear viscosity in central Au+Au collisions  
is estimated  
in the range of incident energies
 3.3 GeV  $\le \sqrt{s_{NN}}\le$ 39 GeV. The simulations are performed  
within a three-fluid model 
employing three different equations of state with and without 
 the   deconfinement  transition. 
In order
to estimate this effective viscosity, 
we consider the  entropy  produced in the 3FD simulations as if it is generated within the 
conventional one-fluid viscous hydrodynamics.
It is found that the effective viscosity within different considered 
scenarios is very similar at the expansion stage of the collision:  
as a function of temperature ($T$) the viscosity-to-entropy ratio behaves as  
$\eta/s \sim 1/T^4$;   
as a function of net-baryon density ($n_B$), 
$\eta/s \sim 1/s$, i.e. it is mainly determined by the density dependence of the entropy density. 
The above dependencies take place along the dynamical trajectories of Au+Au collisions. 
At the final stages of the expansion
the $\eta/s$ values are ranged from $\sim$0.05 at highest considered energies to 
$\sim$0.5 at the lowest ones. 
\PACS{
{25.75.-q}{}, 
\and
{25.75.Nq}{}, 
\and
{24.10.Nz}{} 
}
}

\maketitle


Dissipation in strongly interacting matter is an important property of the produced matter and  
is crucial for understanding the dynamics of heavy-ion collisions. 
Observables that are the most sensitive to the dissipation at the expansion  
stage of the reaction are the elliptic flow and other anisotropic flow coefficients. 
This dissipation deduced from analysis of experimental data at the Large Hadron Collider (LHC) at 
CERN and at top top energies of the Relativistic Heavy Ion Collider
(RHIC) at Brookhaven National Laboratory (BNL) amounts to $\eta/s\approx$ 0.1--0.2 
in terms of the viscosity-to-entropy ratio \cite{Heinz:2013th}. 
The analysis of the STAR data 
in the RHIC Beam Energy Scan (BES) range $\sqrt{s_{NN}} =$ 7.7--200 GeV \cite{Adamczyk:2012ku},
recently performed within a hybrid model \cite{Karpenko:2015xea}, 
indicated that the $\eta/s$ ratio remains approximately in the same range even at lower
BES-RHIC energies. This is definitely in contrast to common expectations that 
at the BES-RHIC energies  the viscosity of the matter
should rapidly rise because the system spends most of its time in the hadronic phase
\cite{Kestin:2008bh}. 

In our recent paper \cite{Ivanov:2014zqa} we found that the model of the 
three-fluid dynamics (3FD) \cite{3FD} equally well describes the STAR data \cite{Adamczyk:2012ku}
on the momentum-integrated elliptic flow of charged particles 
at energies from $\sqrt{s_{NN}} =$ 7.7 to 39 GeV within 
very different scenarios characterized by very different 
equations of state (EoS's)---from a purely hadronic EoS \cite{gasEOS}
to the EoS's involving  deconfinement
 transition \cite{Toneev06}, i.e. a first-order phase transition  
and a smooth crossover one. We assumed that the main reason 
of this good description is that the 
dissipation in the 3FD dynamics with different EoS's is very similar. 
However, we did not present a proof of this assumption 
because the 3FD model does not include viscosity in its formulation.
The dissipation in the 3FD model is present trough
friction interaction between participated fluids rather than a viscosity.
It is highly
difficult to quantitatively express the 3FD dissipation in
terms of the effective viscosity, because this dissipation
depends on dynamics of the collisions rather then
only on parameters of the friction.
In the present paper, in order
to estimate this dissipation in terms of an effective viscosity, 
we consider the  entropy  produced in the 3FD simulations as if it is generated within the 
conventional one-fluid viscous hydrodynamics.


A three-fluid approximation \cite{3FD} is a minimal way to 
simulate a finite stopping power of colliding nuclei at high incident energies.
Within this approximation 
a generally nonequilibrium distribution of baryon-rich
matter is modeled by counter-streaming baryon-rich fluids 
initially associated with constituent nucleons of the projectile
(p) and target (t) nuclei. In addition, newly produced particles,
populating the midrapidity region, are associated with 
a separate net-baryon-free fluid which is called a
``fireball'' fluid (f-fluid).
A certain formation time $\tau_f$ is allowed for the f-fluid, during
which the matter of the fluid propagates without interactions. 
The formation time 
is  associated with a finite time of string formation. 
The physical input
of the present 3FD calculations is described in detail in
Ref.~\cite{Ivanov:2013wha}.

The proper (i.e. in a local rest frame) entropy density of a separate 
fluid ($\alpha$) can be  calculated by means of the thermodynamic relation 
   \begin{eqnarray}
s_{\alpha} = \frac{1}{T_{\alpha}} (\varepsilon_{\alpha} + P_{\alpha} - n_{\alpha} \mu_{\alpha})
   \label{sa}
   \end{eqnarray}
where, $T_{\alpha}$, $\varepsilon_{\alpha}$, $P_{\alpha}$, $n_{\alpha}$ and $\mu_{\alpha}$
are the temperature, the energy density, the pressure, the baryon density 
and the baryon chemical potential of the $\alpha$ fluid.
All these quantities are known from solution of the 3FD equations. 
The total entropy ($S$) is then calculated by integration of the sum of these $s$-densities over volume of the system 
   \begin{eqnarray}
S = \int dV \sum_{\alpha}  u^0_{\alpha}  s_{\alpha}, 
   \label{stot}
   \end{eqnarray}
where $u^0_{\alpha}$ the 0-component of the $\alpha$-fluid 4-velocity, which is introduced 
to transform the proper $s$-density into a common frame of the calculation.

The main idea of estimating an effective viscosity in a nuclear collision consists 
in associating the entropy production within the 3FD simulation with the effect of 
the viscous dissipation within the standard 
viscous hydrodynamics.

In fact, the 3FD dissipation is directly related neither to the shear viscosity nor 
other transport coefficients, i.e. the bulk viscosity ($\zeta$) and thermal conductivity ($\kappa$). 
The dissipation due to these transport coefficients takes place only when gradients
of the collective velocity, temperature and chemical potential exist \cite{Land-Lif,Rischke:1998fq}. 
The 3FD dissipation can, in principal, occur even without any gradients, 
e.g. in two homogeneous counter-streaming media. 
Though the real evolution 
of the nuclear collision gives rise to such gradients. Thus, we can express the 
3FD dissipation in familiar terms by associating it with the shear viscosity. 
The shear viscosity is chosen among other transport coefficients only because 
the dissipation in heavy-ion collisions is traditionally discussed in terms of this quantity.

Under assumption that only the shear viscosity is nonzero among 
the transport coefficient, 
the standard viscous fluid dynamics results in the following equation
for the entropy production \cite{Land-Lif,Rischke:1998fq}
   \begin{eqnarray}
   \label{s-dissipative}
\partial_{\mu} s^{\mu} =  
\frac{1}{T}\pi_{\mu\nu} \partial^{\mu} u^{\nu} 
   \end{eqnarray}
where  $s^{\mu}$, $u^{\mu}$ and $T$ are the entropy four-current, fluid four-velocity and 
temperature, respectively. The stress tensor, $\pi^{\mu\nu}$, reads 
   \begin{eqnarray}
   \label{pi-dissipative}
\pi^{\mu\nu} &=& \eta 
\left(
\partial^{\mu}u^{\nu} + \partial^{\nu}u^{\mu}
-u^{\mu} u_{\lambda}\partial^{\lambda}u^{\nu} 
-u^{\nu} u_{\lambda}\partial^{\lambda}u^{\mu} 
\right)
\cr
&-&
\frac{2}{3} \eta 
\left(g^{\mu\nu} -u^{\mu} u^{\nu} \right)\partial_{\lambda}u^{\lambda}
   \end{eqnarray}
with $\eta$ being the shear viscosity. 
We have to put all other coefficients, i.e. $\zeta$ and $\kappa$, to be zero 
because we can determine only a single quantity from a single equation. 

If the thermal conductivity is zero, the heat flow also vanishes. Then  
we have no other choice but to associate the 
hydrodynamic velocity $u^{\mu}$ is with the baryon flow \cite{Land-Lif} in the 3FD model 
   \begin{eqnarray}
   \label{u-dissipative}
n_B u^{\mu} = n_p u_p^{\mu} + n_t u_t^{\mu}
   \end{eqnarray}
where $n_p$, $n_t$, $u_p^{\mu}$ and $u_t^{\mu}$ are the net-baryon densities and 4-velocities 
of the p- and t-fluids within the 3FD model, respectively, and $n_B$ and $u^{\mu}$ 
those quantities of the unified baryon-rich fluid.  
The mean temperature, $T$, that is also required by Eq. (\ref{s-dissipative}), 
is defined proceeding from common sense, i.e. it is defined
as a local energy-density-weighted temperature
   \begin{eqnarray}
   \label{Tm-dissipative}
T = \sum_\alpha T_{\alpha} \varepsilon_{\alpha} \Big/ \sum_\alpha \varepsilon_{\alpha} .
   \end{eqnarray}
%

Integrating Eq. (\ref{s-dissipative}) over volume, $V$,  we arrive at 
   \begin{eqnarray} 
   \label{eta-s}
\frac{1}{S} \frac{dS}{dt} 
=
\frac{V}{S} \left\langle \frac{1}{T}\pi_{\mu\nu} \partial^{\mu} u^{\nu} \!\right\rangle 
\approx
\frac{\langle\eta\rangle}{\langle s\rangle}
\frac{1}{\langle T\rangle\langle u^{0}\rangle}
\left\langle \frac{1}{\eta}\pi_{\mu\nu} \partial^{\mu} u^{\nu} \!\right\rangle \;\;  
   \end{eqnarray}
where $\langle ...\rangle$ denotes averaging over the volume. Here we also took 
into account that $s^{\mu} = s u^{\mu}$, where $s$ is the proper entropy density, 
and hence $S = V \langle s u^{0}\rangle \approx  V \langle s\rangle  \langle u^{0}\rangle$.  
From this equation, together with definitions (\ref{u-dissipative}) and (\ref{Tm-dissipative}), 
we easily obtain the estimation of the $\eta/s$ ratio. 
In order to facilitate numerical evaluation of terms with time derivatives, 
we also used the approximation 
$\left\langle \pi_{\mu\nu} \partial^{\mu} u^{\nu} \right\rangle
\approx
\left\langle \pi_{\mu\nu}\right\rangle \left\langle\partial^{\mu} u^{\nu} \right\rangle$.  

%
\begin{figure}[bht]
\begin{center}
\includegraphics[width=6.5cm]{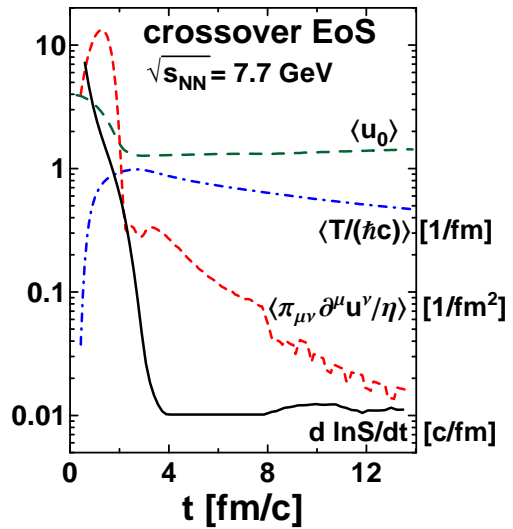}
\end{center}
 \caption{
Time evolution of different factors in Eq. (\ref{eta-s}) at 
 $\sqrt{s_{NN}}=$ 7.7 GeV 
within the crossover scenario. 
}
\label{fig1}
\end{figure}
Figure \ref{fig1} illustrates relative importance of different factors in Eq. (\ref{eta-s}). 
At the initial, highly nonequilibrium stage, when the concept of the viscosity is 
hardly applicable, all the factors reveal fast changes in time. At the late expansion stage 
($t>$ 4 fm/c) only the gradient term 
$\left\langle \pi_{\mu\nu} \partial^{\mu} u^{\nu} \right\rangle$ manifests fast changes, 
the entropy-production factor defines the basic scale of the $\eta/s$ ratio, while the 
two remaining factors, $\langle T\rangle$ and $\langle u^{0}\rangle$, are responsible for 
only relatively insignificant corrections.


The 3FD simulations of central Au+Au collisions  at energies 
3.3 GeV  $\le \sqrt{s_{NN}}\le$ 39 GeV were performed without freeze-out. 
The freeze-out in the 3FD model removes the frozen out matter from the hydrodynamical 
evolution \cite{Russkikh:2006aa,Ivanov:2008zi}. Therefore, in order to keep all the matter in the 
consideration the freeze-out was turned off.

At the initial stage of the reaction, all three fluids coexist in the same 
space-time region, thus describing a certain {\em nonequilibrium} state 
of the matter.  
This short initial stage is followed by a longer stage at which 
the p- and t-fluids are either spatially separated or unified, 
while the f-fluid still overlaps 
with the baryon-rich (p- and t-) fluids to a lesser (at high energies) 
or grater (at lower energies) extent. Therefore, the friction between 
the f-fluid and the baryon-rich fluids still courses the dissipation and 
hence the entropy growth.

%
\begin{figure}[bht]
\begin{center}
\includegraphics[width=7.5cm]{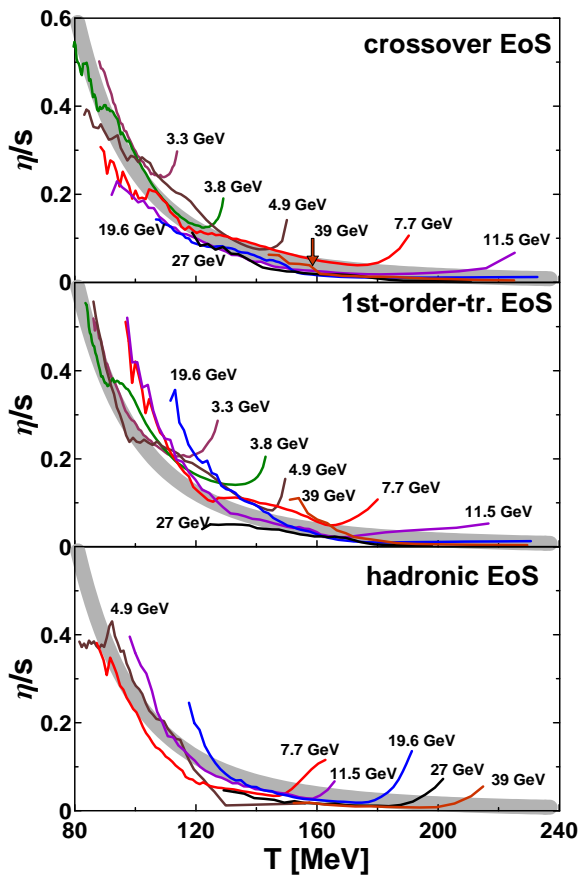}
\end{center}
 \caption{
The  $\eta/s$ ratio as a function of temperature along trajectories of central Au+Au collisions at 
various collision energies $\sqrt{s_{NN}}$ 
within different scenarios. 
The gray band in all the panels is the function $(T_0/T)^4$, where $T_0=$ 71 MeV. 
}
\label{fig4}
\end{figure}
%

%
\begin{figure}[tbh]
\begin{center}
\includegraphics[width=7.5cm]{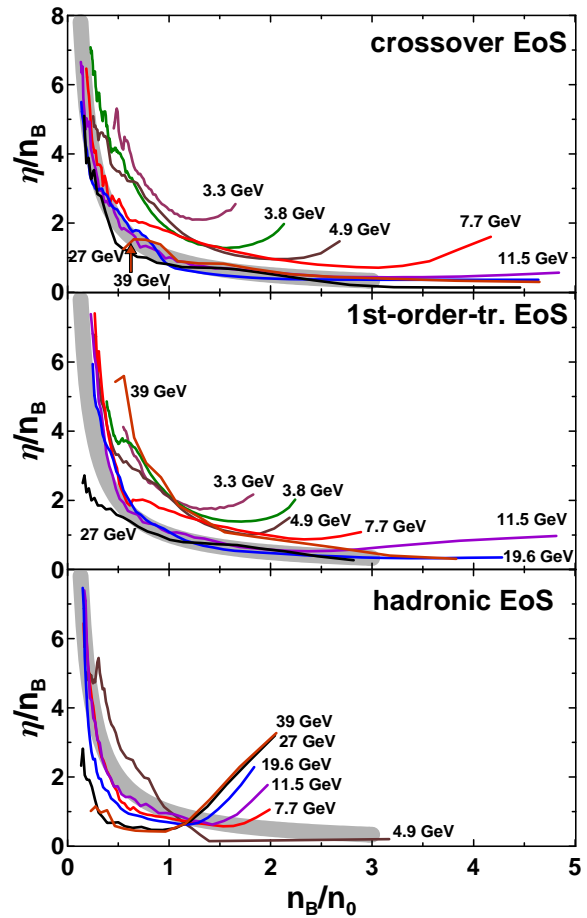}
\end{center}
 \caption{
The  kinematic viscosity, $\eta/n_B$, as a function of net-baryon density, $n_B$, 
along trajectories of central Au+Au collisions at 
various collision energies $\sqrt{s_{NN}}$ 
within different scenarios. 
The gray band in all the panels is the function $n_0/n_B$, where $n_0=$ 0.15 fm$^{-3}$ 
is the normal nuclear density. 
}
\label{fig7}
\end{figure}

In Fig. \ref{fig4} the results on the $\eta/s$ ratio are presented   
as a functions of the mean temperature $\langle T\rangle$ 
[cf. Eqs.  (\ref{u-dissipative})-(\ref{eta-s})] along dynamical trajectories of central Au+Au collisions at 
various collision energies $\sqrt{s_{NN}}$ within different scenarios based on  
a purely hadronic EoS \cite{gasEOS} and those involving the   deconfinement
 transition \cite{Toneev06}, i.e. a first-order phase transition  
and a smooth crossover one. 
We plot the $\eta/s$ ratio as a function of the temperature rather than time 
because the temperature is a natural argument of the $\eta/s$ quantity. 
The mean temperature is also a function of time. 
The viscosity is meaningful when nonequilibrium is weak. 
Therefore, it should be analyzed at the expansion stage of the collision following 
the fast highly nonequilibrium stage. 
In terms of the $\eta/s$ ratio of Fig. \ref{fig4}, 
the expansion stage takes place at lower temperatures up to the minimum of 
the $\eta/s$ ratio.  
The $\eta/s$ curves are continued to higher temperatures after the minimum 
only for the sake of convenience of labeling them.

The results that manifest fluctuations exceeding the 
the scale of the plot are omitted.   
These fluctuations are a consequence of 
the numerical calculation of the derivatives that courses 
a loss of accuracy. In view of this low accuracy and a very approximate nature of 
Eq. (\ref{eta-s}) itself, the present results on the $\eta/s$ ratio should be considered as an 
order-of-magnitude estimation. For the sake of the graphic representation, we apply running 
average procedure to the results of the direct calculation in such a way that 
the $\eta/s$ ratio is averaged over each sequential five time steps. Though these running-average results 
are not completely smooth, they exhibit much weaker fluctuations.

At high temperatures $T\gsim$ 160 GeV in collisions with $\sqrt{s_{NN}}>$ 10 GeV, the  $\eta/s$ ratio turns out to 
be noticeably smaller that the conjectured  lowest bound for this quantity $1/(4\pi)$ 
\cite{Kovtun:2004de}. 
Even in view of the above discussed roughness of the present estimate, 
that small values $\eta/s$ should be attributed to the 3FD model itself. 
This is certainly a theoretical shortcoming of the model. 
At the final stages of the expansion%
\footnote{We avoid the term of freeze-out because the freeze-out within the 3FD model is an 
extended in time process which continues over the whole expansion stage \cite{Russkikh:2006aa,Ivanov:2008zi}.}
the $\eta/s$ values are ranged from $\sim$0.05 at highest considered energies to 
$\sim$0.5 at the lowest ones.

We further focus on qualitative properties 
of the deduced $\eta/s$ ratio. 
As seen from Fig. \ref{fig4}, the temperature dependence of the $\eta/s$ ratio 
at the expansion stages of collisions at various collision energies
is very similar within different scenarios. This dependence is approximately described 
by $1/T^4$ low, i.e. this ratio decreases with the temperature rise, as it is commonly expected. 
It is important to emphasize that this is the $T$-dependence along dynamical trajectories 
of collisions, along which the mean net-baryon density, $n_B$, also changes. 
The density dependence of the $\eta/s$ ratio is also very similar within different scenarios, 
though it does not follow any universal low in terms of $n_B$. 
In the case of another representation,  
i.e. in terms of the kinematic viscosity $\eta/n_B$, it is more spectacular, as it is seen 
from Fig. \ref{fig7}. The $n_B$-dependence of the kinematic viscosity along dynamical trajectories 
of collisions approximately follows the low of $1/n_B$. 
Of course, different numeric factors are required for different collision energies: 
from $\sim 2n_0$ at low $\sqrt{s_{NN}}$ to $\sim 0.5 n_0$ at highest considered $\sqrt{s_{NN}}$, 
where $n_0=$ 0.15 fm$^{-3}$ is the normal nuclear density. 
Thus, we get $\eta/s \sim 1/s$ in terms of the $\eta/s$ ratio. 
This is in agreement with the result of Ref. \cite{Khvorostukhin:2010aj,Denicol:2013nua}, 
where it was found that the reduction of the $\eta/s$ ratio
with $n_B$ rise happens mostly because of $s$ increase.

Apparently, the similarity of the deduced effective viscosity within different scenarios 
is the main reason why all considered scenarios equally well 
reproduce the measured integrated elliptic flow  of charged particles \cite{Ivanov:2014zqa}. 
In this respect, the estimated $\eta/s$ can be considered as that deduced from 
experimental data \cite{Adamczyk:2012ku} by means of the 3FD analysis. 
Of course, this $\eta/s$ is model dependent and is not unique.


In conclusion, we estimated the effective viscosity in central Au+Au collisions at collision energies
from $\sqrt{s_{NN}}=$ 3.3 GeV to 39 GeV within different scenarios in order to quantify 
the dissipation in the 3FD model. 
To estimate this dissipation in therms of the effective shear viscosity (more precisely, 
the $\eta/s$ ratio), we considered the entropy produced in the 3FD model 
as if it was generated within the conventional one-fluid viscous hydrodynamics.

It is found that the effective viscosity within different considered 
scenarios (with and without deconfinement transition) 
is very similar at the expansion stage of the collision:  
as a function of tempereture ($T$), 
$\eta/s \sim 1/T^4$ and quantitatively is very similar within different 
scenarios;   
as a function of net-baryon density ($n_B$), 
$\eta/s \sim 1/s$, i.e. is mainly determined by the density dependence of the entropy density. 
The above dependencies take place along the dynamical trajectories of Au+Au collisions. 
In the hadronic scenario, the reported small values of the  $\eta/s$ ratio 
at high collision energies, $\sqrt{s_{NN}}>$ 10 GeV, were 
achieved due to an artificial enhancement \cite{3FD,Ivanov:2013wha} of the friction forces 
estimated on the basis of experimental proton-proton cross sections \cite{Sat90}. 
This enhancement was required to reproduce the observed baryon stopping at high energies. 

At the final stages of the expansion
the $\eta/s$ values are ranged from $\sim$0.05 at highest considered energies to 
$\sim$0.5 at the lowest ones. This result does not contradict the finding of Ref. 
\cite{Karpenko:2015xea}, where average $\eta/s$ over the expansion stage values were reported, 
because in our case the  $\eta/s$ ratio turns out to be strongly temperature dependent.


Fruitful discussions with D.N. Voskresensky 
are gratefully acknowledged.
The calculations were performed at the computer cluster of GSI (Darmstadt). 
This work was partially supported  by  grant NS-932.2014.2.

\end{document}